\documentclass[a4paper,11pt]{article}
\usepackage{amssymb,amsmath,bm}  % for math
\usepackage{graphics,graphicx}   % for figures
\usepackage{array,booktabs}      % for tables
\usepackage{authblk}  % for footnote style author/affiliation
\usepackage{slashed}

\usepackage{cite}
\usepackage{xcolor}
\usepackage[margin=2.5cm]{geometry}

\DeclareUnicodeCharacter{2212}{-}
\usepackage[linktocpage,
colorlinks = true,
linkcolor = blue,
urlcolor  = blue,
citecolor = red,
anchorcolor = blue]{hyperref}

\usepackage{epsfig}

\numberwithin{equation}{section}

\numberwithin{equation}{section}

\title{\textbf{Gravitational wave effects and phenomenology of a
		two-component dark matter model }}
\author[a]{Mojtaba Hosseini \thanks{mojtaba\textunderscore hosseini@semnan.ac.ir }}
\author[a]{Seyed Yaser Ayazi\thanks{syaser.ayazi@semnan.ac.ir}}
\author[b]{Ahmad Mohamadnejad\thanks{mohamadnejad.a@lu.ac.ir}}
\affil[a]{Physics Department, Semnan University, P.O. Box. 35131-19111, Semnan, Iran}
\affil[b]{Department of Physics, Lorestan University, Khorramabad, Iran}

\date{\today}
\begin{document}

\baselineskip 0.6 cm
\maketitle

\begin{abstract}

We study an extension of the Standard Model (SM) which could have two candidates
for dark matter (DM) including a Dirac fermion and a vector dark matter (VDM) under a new $U(1)$ gauge group in the hidden sector. The model is classically scale- invariant
and the electroweak symmetry breaks because of loop effects. We investigate the
 parameter space allowed by current experimental constraints and phenomenological bounds. We probe the parameter space of the model in the mass range  $1< M_V<5000$ GeV and $1<M_{\psi}<5000$ GeV. It has been shown that there are many points in this mass range that are in agreement with all phenomenological constraints. The electroweak phase transition has been discussed and it has been shown that there is region in the parameter space of the model consistent
with DM relic density and direct detection constraints that, at the same time, can lead to first order electroweak phase transition. The gravitational waves produced during the phase transition could be probed by future space-based interferometers such as LISA and BBO.

\end{abstract}

\section{Introduction} \label{sec1}
The Standard Model (SM) has been the most effective way to describe the functioning of the world around us, but is incomplete because of challenges like matter–antimatter asymmetry, hierarchy problem, and dark matter (DM). DM is estimated to make up approximately 27 \% of our universe, as indicated by a lot of astrophysical and cosmological evidence\cite{Bertone:2016nfn}. One of the main goals of particle physicists is to predict and find a particle that can satisfy the properties of DM, which can be a window to physics beyond the standard model (BSM).

Weakly interacting massive particles (WIMPs) are the most popular candidate for DM, with the freeze-out scenario being the most popular choice\cite{Feng:2022rxt}. WIMP paradigm is essential background for almost any discussion of particle DM and the triple coincidence of motivations from particle theory, particle experiment, and cosmology is known as the WIMP miracle. However, no trace of DM has so far been found in direct detection experiments. Due to strong constraints on direct detection experiments in one-component DM models, multi-component DM models seem more appropriate in some ways
\cite{Zurek:2008qg,Profumo:2009tb,Aoki:2012ub,Biswas:2013nn,Gu:2013iy,Aoki:2013gzs,Kajiyama:2013rla,Bian:2013wna,Bhattacharya:2013hva,Geng:2013nda,Esch:2014jpa,Dienes:2014via,Bian:2014cja,Geng:2014dea,DiFranzo:2016uzc,Aoki:2016glu,DuttaBanik:2016jzv,Pandey:2017quk,Borah:2017xgm,Herrero-Garcia:2017vrl,Ahmed:2017dbb,PeymanZakeri:2018zaa,Aoki:2018gjf,Chakraborti:2018lso,Bernal:2018aon,Poulin:2018kap,Herrero-Garcia:2018qnz,YaserAyazi:2018lrv,Elahi:2019jeo,Borah:2019epq,Bhattacharya:2019fgs,Biswas:2019ygr,Nanda:2019nqy,Yaguna:2019cvp,Belanger:2020hyh,VanDong:2020bkg,Khalil:2020syr,DuttaBanik:2020jrj,Hernandez-Sanchez:2020aop,Chakrabarty:2021kmr,Yaguna:2021vhb,DiazSaez:2021pmg,DiazSaez:2021pfw,Kim:2022sfg,Ho:2022erb,Costa:2022oaa,Bhattacharya:2022qck,Bhattacharya:2022wtr,Bhattacharya:2018cgx,Bhattacharya:2016ysw,Mohamadnejad:2021tke}.

In the SM, electroweak phase transition is of the second order\cite{Kajantie:1996mn,Aoki:1999fi} and does not generate the gravitational wave (GW) signal (for a recent review see \cite{Weir:2017wfa}). A first-order phase transition can be caused by certain extensions of the SM and the DM candidate, leading to the creation of GWs\cite{Chala:2016ykx,Soni:2016yes,Flauger:2017ged,Baldes:2017rcu,Chao:2017vrq,Beniwal:2017eik,Huang:2017rzf,Huang:2017kzu,Madge:2018gfl,Bian:2018mkl,Bian:2018bxr,Shajiee:2018jdq,Kannike:2019wsn,YaserAyazi:2019caf,Mohamadnejad:2019vzg,Kannike:2019mzk,Paul:2019pgt,Barman:2019oda,Marfatia:2020bcs,Alanne:2020jwx,Han:2020ekm,Wang:2020wrk,Deng:2020dnf,Chao:2020adk,Zhang:2021alu,Liu:2021mhn,Hashino:2018zsi,Athron:2023xlk,Khoze:2022nyt}.
In the early universe when two local minima of free energy (potential) co-exist for some range of temperatures (critical temperature), strongly first-order electroweak phase transition can take place. After that, the relevant scalar fields can quantum-mechanically tunnel into the new phase and through the nucleation of bubbles and collide with each other to cause a significant background of GWs\cite{Witten:1980ez,Guth:1981uk,Steinhardt:1980wx,Steinhardt:1981ct,Witten:1984rs}.The discovery of GWs resulting from the first-order phase transition can be the consequence of physics BSM, which can be a supplement to ground experiments like those conducted using the Large Hadron Collider (LHC). Unlike GWs from strong astrophysical sources\cite{LIGOScientific:2016aoc}, these waves have a range between millihertz and decihertz\cite{Caprini:2019egz}. The Laser Interferometer Space Antenna (LISA)\cite{LISA:2017pwj} and  Big Bang Observer (BBO)\cite{Crowder:2005nr} are two space-based GW detectors which are expected to observe GWs resulting from cosmological phase transitions in future years. On the other hand, one of the Sakharov conditions\cite{Shaposhnikov:1987tw} which explains the matter–antimatter asymmetry in universe is the thermal imbalance that occurs in first-order phase transitions.

 As mentioned, one of the fundamental challenges of particle physics  is the hierarchy problem. A potential solution to this problem is to drop the Higgs mass term in the potential.  SM without the Higgs mass term is scale-invariant. In this paper, we present a classically scale-invariant extension of the SM where all the particle masses are generated using the  Coleman-Weinberg mechanism\cite{Coleman:1973jx}. The model includes three new fields: a fermion, a complex singlet scalar and a vector field with $U_D(1)$ gauge symmetry. We examine two scenarios in this paper. In scenario A, we consider only the fermionic field as DM. In scenario B, both fermionic and vector fields are considered as DM. We probe the parameter space of the model according to constraints from relic density and direct detection. DM relic density is reported by the Planck Collaboration\cite{Planck:2018vyg} and DM-Nucleon cross section is constrained by XENONnT experiment results\cite{XENON:2023sxq}. We investigate the possibility of the electroweak phase transition with respect to the bounded parameter space, where we use the effects of the effective potential of the finite temperature. We probe the parameter space of the model which is consistent with the said phenomenological constraints and also leads to a strong first-order electroweak phase transition. The GW signal resulting from this phase transition also has been studied in the LISA and BBO detectors.

Here is the organization of the paper. In the next section, we introduce the model. In Sect.~\ref{Scenario a}, we study the phenomenology of the Scenario A including relic density, direct detection, invisible Higgs decay and the resulting GWs. Section~\ref{Scenario b} is dedicated to the phenomenology of the Scenario B and its GW signals. Finally, our conclusion comes in Sect.~\ref{Conclusion}.

\section{The model} \label{sec2}
In this section, we consider an extension of the SM to explain DM phenomenology. In this regard, the model contains three new fields in which a vector dark matter $V_{\mu}$ and a Dirac fermion field $\psi$ can play the role of DM. A complex scalar,$S$, mediates between SM and the dark sector. In the model $V_{\mu}$, $\psi$ and $S$ are charged under a new dark $U(1)_D$ gauge group. All of these fields are singlet under SM gauge groups. We suppose the mass of the fermion was produced by breaking of dark $U_D(1)$ gauge symmetry, and so was constrained by other parameters of the model.
However, in the model, the dark sector is invariant under the transformations of the $U_D(1)$ gauge group:
\begin{align} \label{2-1}
& \psi_L \rightarrow e^{i Q_l \alpha(x)} \psi_L, \nonumber \\
& \psi_R \rightarrow e^{i Q_r \alpha(x)} \psi_R, \nonumber \\
& S \rightarrow e^{i Q_s \alpha(x)} S, \nonumber \\
& V_{\mu} \rightarrow V_{\mu} - \frac{1}{g_v} \partial_{\mu}{\alpha(x)} .
\end{align}
where the $U(1)_D$ charge of the new particles, $ Q_{l,r,s} $, are given in Table \ref{Table}.

The Lagrangian for the model is given by the following
renormalizable interactions,
\begin{align}
{\cal L} ={\cal L}_{SM}+ i\bar\psi_L  \gamma^{\mu}D_{\mu}\psi_L+ i \bar\psi_R \gamma^{\mu}D_{\mu}\psi_R
-g_s \bar\psi_L \psi_R S+h.c.-\frac{1}{4} V_{\mu \nu} V^{\mu \nu}+ (D_{\mu} S)^{*} (D^{\mu} S)- V(H,S).
\label{eq:lagrangian}
\end{align}
where $ {\cal L} _{SM} $ is the SM Lagrangian without the Higgs potential term, The covariant derivative is
\begin{align}
& D_{\mu}= (\partial_{\mu} + i Q g_{v} V_{\mu}), \quad \text{and}\nonumber \\
& V_{\mu \nu}= \partial_{\mu} V_{\nu} - \partial_{\nu} V_{\mu},\end{align}

On the imposition of dark charge conjugation symmetry,
we do not assume a kinetic mixing term between $V_{\mu}$ and the $U(1)_Y$ gauge boson of the SM.

\begin{table}
\begin{center}
\begin{tabular}{| l | l | l | l |}
\hline
Field&$S$&$\psi_L$&$\psi_R$\\ \hline
$U(1)_D$ charge&1&$\frac{1}{2}$&$-\frac{1}{2}$\\ \hline
\end{tabular}
\end{center}
\caption{\label{Table}The charges of the dark sector particles under the new $U(1)_D$ symmetry.}
\end{table}

The most general scale-invariant potential $ V(H,S) $ that is renormalizable and invariant
under gauge symmetry is
\begin{equation}
V(H,S) = \frac{1}{6} \lambda_{H} (H^{\dagger}H)^{2} + \frac{1}{6} \lambda_{S} (S^{*}S)^{2} + 2 \lambda_{S H} (S^{*}S) (H^{\dagger}H). \label{2-3}
\end{equation}
Note that the quartic portal interaction, $ \lambda_{S H} (S^{*}S) (H^{\dagger}H) $, is the only connection between the dark sector and the SM.

SM Higgs field $ H $ as well as dark scalar $S$ can receive VEVs, respectively breaking the electroweak and $U_{D}(1)$ symmetries.
In unitary gauge, the imaginary component of $ S $ can be absorbed as the longitudinal part of $V_{\mu}$.
In this gauge, we can write
\begin{equation}
H = \frac{1}{\sqrt{2}} \begin{pmatrix}
0 \\ h_{1} \end{pmatrix} \, \, \, and \, \, \, S = \frac{1}{\sqrt{2}} h_{2} , \label{2-4}
\end{equation}
where $ h_{1} $ and $ h_{2} $ are real scalar fields which can receive VEVs. Now, the tree-level potential becomes
\begin{equation}
V^{tree} = \frac{1}{4 !} \lambda_{H} h_{1}^{4} + \frac{1}{4 !} \lambda_{S} h_{2}^{4} + \frac{1}{2} \lambda_{S H} h_{1}^{2} h_{2}^{2}. \label{2-5}
\end{equation}
There is a $Z_2$ symmetry for $\psi$, making it a stable particle. In addition, if the mass of $V_{\mu}$ is less than two times of the mass of $\psi$, then both $V_{\mu}$ and $\psi$ are viable DM candidates.

For the Hessian matrix, we define:
\begin{equation}
H_{ij} (h_{1},h_{2}) \equiv \frac{\partial^{2} V^{tree}}{\partial h_{i} \partial h_{j}}. \label{2-6}
\end{equation}
Necessary and sufficient conditions for local minimum of $ V^{tree} $ in which  vacuum expectation values $ \langle h_{1} \rangle = \nu_{1} $ $\rm and$ $ \langle h_{2} \rangle = \nu_{2} $, have be written as:
\begin{align}
& \frac{\partial V^{tree}}{\partial h_{i}} \bigg\rvert_{\nu_{1},\nu_{2}} = 0 \label{2-7} \\
& \frac{\partial^{2} V^{tree}}{\partial h_{i} ^{2}} \bigg\rvert_{\nu_{1},\nu_{2}} > 0  \label{2-8} \\
& det(H (\nu_{1},\nu_{2})) > 0 , \label{2-9}
\end{align}
where $ det(H (\nu_{1},\nu_{2})) $ is determinant of the Hessian matrix.
Condition (\ref{2-7}) for non-vanishing VEVs leads to the following constraints
\begin{align}
&\lambda_{H} \lambda_{S} = (3! \lambda_{S H})^{2}
&\frac{\nu_{1}^{2}}{\nu_{2}^{2}} = - \frac{3! \lambda_{S H}}{\lambda_{H}}. \label{2-10}
\end{align}
Conditions (\ref{2-7}) and (\ref{2-8}) require $ \lambda_{H} > 0 $, $ \lambda_{S} > 0 $, and $ \lambda_{S H} < 0 $. However, condition (\ref{2-9}) will not be satisfied, because $ det(H (\nu_{1},\nu_{2})) = 0 $. When the determinant of the Hessian matrix is zero, the second derivative test is inconclusive, and the point $ (\nu_{1},\nu_{2}) $ could be any of a minimum, maximum or saddle point. However, in the model, constraint (\ref{2-10}) is defined as a flat direction, in which $ V^{tree} = 0 $. Thus, it is the stationary line or a local minimum.

The important point is that for other directions $ V_{eff}^{1-loop(T=0)} > 0 $, and the tree level potential only vanishes along the flat direction. Thus, the full potential of the theory will be dominated by higher-loop contributions along flat direction and specifically by the one-loop effective potential. Considering one-loop effective potential, $ V_{eff}^{1-loop} $, can lead to a small curvature in the flat direction which picks out a specific value along the ray as the minimum with $ V_{eff}^{1-loop} < 0 $ and vacuum expectation value $ \nu^{2} = \nu_{1}^{2} + \nu_{2}^{2} $ characterized by a RG scale $ \Lambda $. Since at the minimum of the one-loop effective potential $ V^{tree} \geqslant 0 $ and $ V_{eff}^{1-loop} < 0 $, the minimum of $ V_{eff}^{1-loop} $ along the flat direction (where $ V^{tree}=0 $) is a global minimum of the full potential, and so spontaneous symmetry breaking takes place. As a result, we suppose $ h_{1} \rightarrow \nu_{1} + h_{1} $ and $ h_{2} \rightarrow \nu_{2} + h_{2} $ and the electroweak symmetry breaks with  value $ \nu_{1} = 246 $ GeV.
In tree level potential since $ h_{1} $ and $ h_{2} $ mix with each other, they can be rewritten by the mass eigenstates $ H_{1} $ and $ H_{2} $ as
\begin{equation}
\begin{pmatrix}
H_{1}\\H_{2}\end{pmatrix}
=\begin{pmatrix} cos \alpha~~~  -sin \alpha \\sin \alpha  ~~~~~cos \alpha
\end{pmatrix}\begin{pmatrix}
h_1 \\  h_{2}
\end{pmatrix}, \label{2-11}
\end{equation}
where $ H_{2} $ is along the flat direction, thus $ M_{H_{2}} = 0 $, and $ H_{1} $ is perpendicular to the flat direction which we identify it as the SM-like Higgs observed at the LHC with $ M_{H_{1}} = 125 $ GeV. After the symmetry breaking, we have the following constraints:
\begin{align}
& \nu_{2} =  \frac{M_{V}}{g_v} , &\nonumber
& sin \alpha =  \frac{\nu_{1}}{\sqrt{\nu_{1}^{2}+\nu_{2}^{2}}} \nonumber \\
&M_{\psi}= \frac{g_sM_{V}}{\sqrt{2}g_v}&\nonumber
& \lambda_{H} =  \frac{3 M_{H_{1}}^{2}}{ \nu_{1}^{2}} cos^{2} \alpha  \nonumber  \\
& \lambda_{S} =  \frac{3 M_{H_{1}}^{2}}{ \nu_{2}^{2}} sin^{2} \alpha  &\nonumber
& \lambda_{S H} =  - \frac{ M_{H_{1}}^{2}}{2 \nu_{1} \nu_{2} } sin \alpha \, cos \alpha , \\ \label{2-12}
\end{align}
where $M_{\psi}$ and $M_{V}$ are the masses of vector and fermion fields after symmetry breaking. Conditions (\ref{2-12}) constrain free parameters of the model up to three independent parameters.
We choose $M_{V}$,  $M_{\psi} $ and  $g_v$ as the independent parameters of the model.

Since in tree level, $M_{H_{2}}=0$,  and the elastic scattering cross section of DM off nuclei becomes severely large, the model actually is excluded by direct detection experiments. However, the radiative corrections give a mass to the massless eigenstate $ H_{2} $. One-loop corrections to the potential, via the Gildener–Weinberg formalism \cite{Gildener:1976ih}, shift the scalon mass to
the values that can be even higher than SM Higgs mass. Along the flat direction, the one-loop effective potential has the general form \cite{Gildener:1976ih}
\begin{equation}
V_{eff}^{1-loop(T=0)} = a H_{2}^{4} + b H_{2}^{4} \, \ln \frac{H_{2}^{2}}{\Lambda^{2}}  , \label{2-13}
\end{equation}
where  $ a $ and $ b $ are the dimensionless constants that given by
\begin{align}
& a =  \frac{1}{64 \pi^{2} \nu^{4}}  \sum_{k=1}^{n} g_{k}  M_{k}^{4} \ln \frac{M_{k}^{2}}{\nu^{2}}  , \nonumber \\
& b = \frac{1}{64 \pi^{2} \nu^{4}} \sum_{k=1}^{n} g_{k}  M_{k}^{4} . \label{2-14}
\end{align}
In (\ref{2-14}), $ M_{k} $ and $ g_{k} $ are the  tree-level mass and the internal degrees of freedom of the particle $ k $, respectively (In our convention $ g_{k} $
takes positive values for bosons and negative values for
fermions).

Minimizing (\ref{2-13}) shows that the potential has a non-trivial stationary point at a value of the RG scale $ \Lambda $, given by
\begin{equation}
\Lambda = \nu \exp \left( \frac{a}{2b} + \frac{1}{4} \right)   , \label{2-15}
\end{equation}
Eq.~(\ref{2-15}) can now be used to find the form of the one-loop effective potential along the flat direction in terms of the one-loop VEV $ \nu $
\begin{equation}
V_{eff}^{1-loop(T=0)} = b H_{2}^{4} \, \left( \ln \frac{H_{2}^{2}}{\nu^{2}} - \frac{1}{2} \right) , \label{2-16}
\end{equation}
It is remarkable that the scalon does not remain massless beyond the tree approximation. Regarding $ V_{eff}^{1-loop(T=0)} $, $ M_{H_{2}} $  will be
\begin{equation}
M_{H_{2}}^{2} = \frac{d^2 V_{eff}^{1-loop(T=0)}}{d H_{2}^{2}} \bigg\rvert_{\nu} = 8 b \nu^{2} . \label{2-17}
\end{equation}
Considering (\ref{2-14}),  the scalon mass can be expressed in terms of other particle masses
\begin{equation}
M_{H_{2}}^{2} = \frac{1}{8 \pi^{2} \nu^{2}} \left( M_{H_{1}}^{4} + 6  M_{W}^{4} + 3  M_{Z}^{4} + 3  M_{V}^{4} - 12 M_{t}^{4} -4M_{\psi}^{4}  \right) . \label{2-18}
\end{equation}
where $ M_{W,Z,t} $ are the masses of W, Z gauge bosons, and top quark, respectively. As mentioned before, $ M_{H_{1}} = 125 $ GeV and $\nu^{2} = \nu_{1}^{2} + \nu_{2}^{2} $.
Notice that in order for $ V_{eff}^{1-loop(T=0)} $ be a
minimum, it must be less than the value of the potential at the origin, hence it must be negative. From (\ref{2-18}), we have a constraint on the parameter space of our model where  $ M_{H_{2}}>0$.

Note that according to (\ref{2-18}) and (\ref{2-12}), $ M_{H_{2}} $ is completely determined by the independent parameters of the model, i.e., $ M_{V} $, $M_{\psi}$ and the coupling $ g_v $ . These constraints are due to the scale invariance conditions which were imposed on the model. Depending on the new particle masses of our model, we examine two different scenarios. In scenario A, the $\psi$ field is considered as DM. In scenario B, $M_V< 2M_{\psi} $ and both $V_{\mu}$ and $\psi$ fields are considered as DM. In the following, we examine the phenomenology of each scenario separately.

\section{Scenario A}\label{Scenario a}

\subsection{DM phenomenology}
\subsubsection{Relic density}

\begin{figure}
	\begin{center}
		\centerline{\hspace{0cm}\epsfig{figure=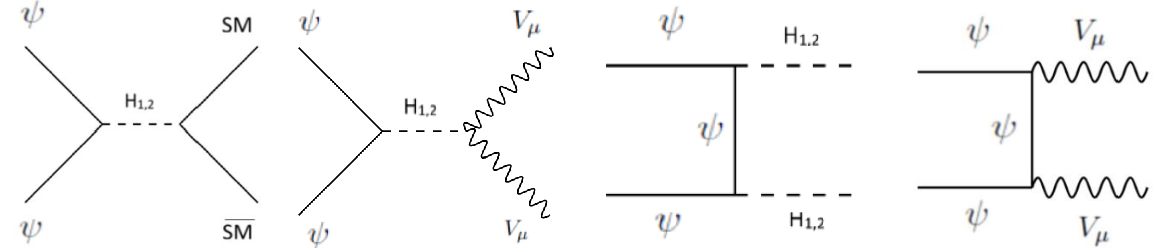,width=12cm}}
		\centerline{\vspace{-0.2cm}}
		\caption{The relevant Feynman diagrams for DM relic density production cross section.} \label{feynman2}
	\end{center}
\end{figure}

\begin{figure}%[!htb]
	\begin{center}
		\centerline{\hspace{0cm}\epsfig{figure=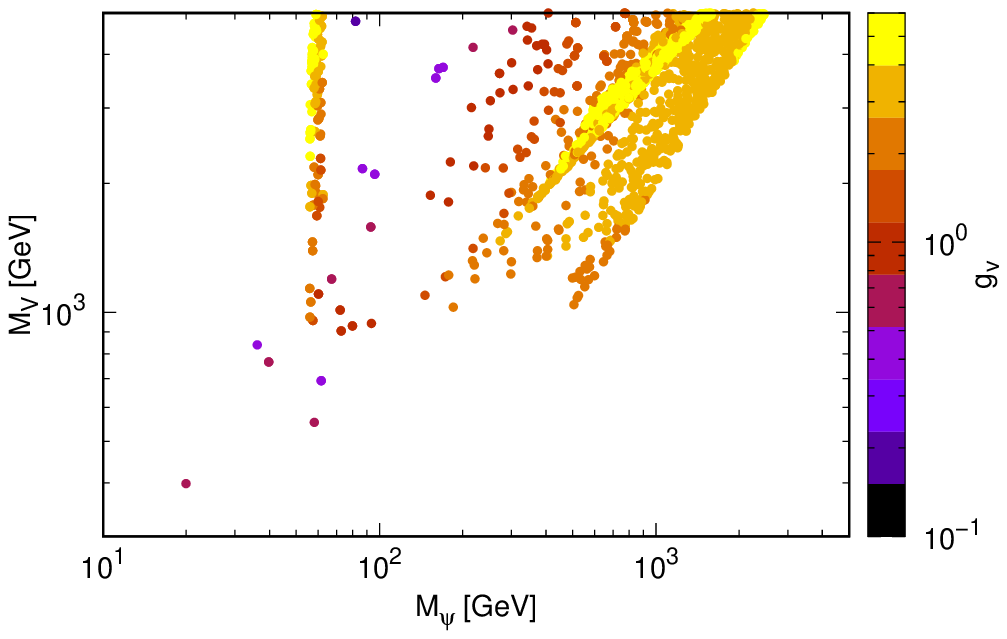,width=12cm}}
		\centerline{\vspace{-0.2cm}}
		\caption{The allowed range of parameter space consistent with DM relic density.} \label{Relic2}
	\end{center}
\end{figure}

In this case, since we do not assume $M_V<2M_{\psi}$, the only candidate for DM is $\psi$. In the WIMP scenario, at first the early universe is hot and very dense and all particles are in thermal equilibrium. Then the universe cools until its temperature falls below the mass of DM particles, and the amount of DM becomes Boltzmann suppressed, dropping exponentially as $e^{-m_X/T}$. As the universe expands, the DM particles are diluted and can no longer find each other until they are annihilated and are out of thermal equilibrium with the SM particles. Then the DM particles freeze out and their number asymptotically reaches a constant value as their thermal relic density. The evolution of the number density of DM particle $(n_{\psi})$ with time is governed by the Boltzmann equation:
\begin{equation} \label{44}
\dot{n_{\psi}} + 3Hn_{\psi} = -\langle\sigma_{ann} \nu_{rel}\rangle [n_{\psi} ^2 -(n_{\psi} ^{eq})^2] ,
\end{equation}
where $H$ is the Hubble parameter and $n_{\psi} ^{eq} \sim (m_{\psi} T)^{3/2} e^{-m_{\psi} /T} $ is the particle density before particles get out of equilibrium. The relevant Feynman diagrams for DM production are shown in  the Fig.\ref{feynman2}. We calculate the relic density
numerically for the $\psi$ particle by implementing  the model into micrOMEGAs \cite{Barducci:2016pcb}. Figure \ref{Relic2} shows the parameter space of the model in agreement with the observed density relic\cite{Planck:2018vyg}. As can be seen, there is agreement for $400< M_V<5000$ GeV , $20<M_{\psi}<2500$ GeV and $0.1<  g_v <6$.

\subsubsection{Direct detection}

WIMPs may be detected by the scattering off normal matter through processes
$X SM$$\rightarrow$$X SM$. Given a typical WIMP mass of $m_X \sim 100$ GeV and WIMP velocity $\upsilon \sim 10^{-3}$, the deposited recoil energy is limited to $\sim 100$ keV, so detection requires highly-sensitive, low-background and deep-site detectors. Such detectors are insensitive to very strongly-interacting
DM, which would be stopped in the atmosphere or earth and would be undetectable
underground. The spin-independent direct detection(DD) cross sections of  ${\psi}$ were obtained using the micrOMEGAs package \cite{Barducci:2016pcb}.

\begin{figure}%[!htb]
	\begin{center}
		\centerline{\hspace{0cm}\epsfig{figure=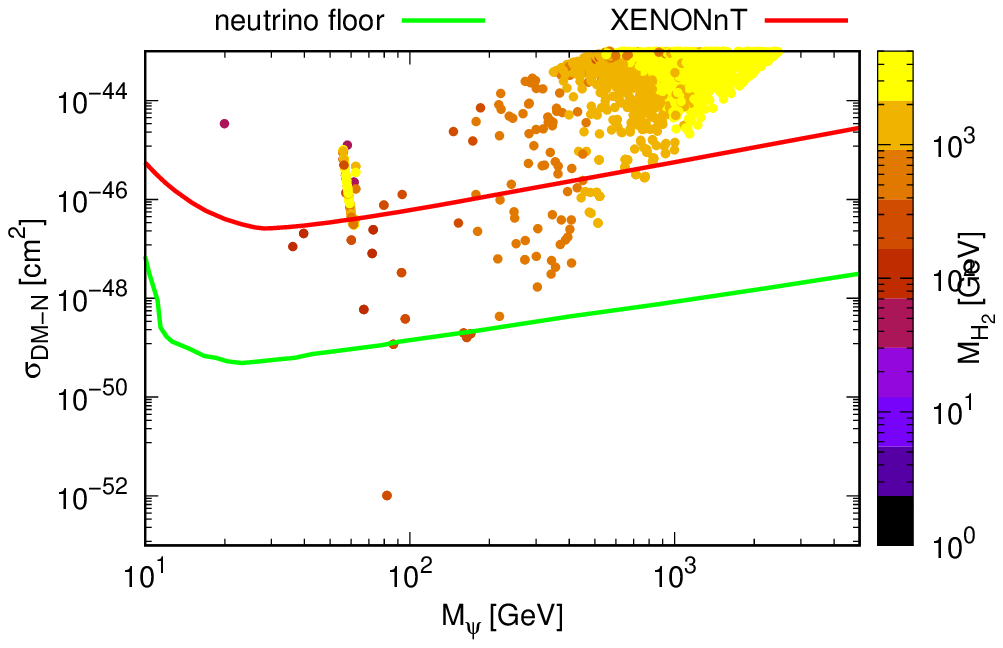,width=12cm}}
		\centerline{\vspace{-0.2cm}}
		\caption{The allowed range of parameter space consistent with DM relic density and DD.} \label{direct detection 2}
	\end{center}
\end{figure}

 In Fig~\ref{direct detection 2}, the parameter space of the model is drawn in agreement with the limits of relic density, XENONnT and neutrino floor. For  $20<M_{\psi}<1000$ GeV, there will be points of the parameter space that fall below the XENONnT limit. As can be seen from Fig.~\ref{direct detection 2}, for $M_{\psi}<M_{H_1}/2$ there will be points that allow the investigation of the invisibility Higgs decay, which will be investigated in the next section.

\subsubsection{Invisible Higgs decay}
As mentioned, for a parameter space consistent with $\psi$ relic density and DD, SM Higgs($H_1$) can only kinematically decay into a pair of ${\psi}$. Therefore, $H_1$ can contribute to the invisible decay mode with a branching ratio:
\begin{eqnarray}
Br(H_1\rightarrow \rm Invisible)& =\frac{\Gamma(H_1\rightarrow {\psi}{\psi})}{\Gamma(h)_{SM}+\Gamma(H_1\rightarrow {\psi}{\psi})},
\label{decayinv1}
\end{eqnarray}

where $\Gamma(h)_{SM}=4.15 ~ \rm [MeV]$ is total width of Higgs boson\cite{LHCHiggsCrossSectionWorkingGroup:2011wcg}. The partial width for process $H_1\rightarrow {\psi}{\psi}$ is given by:
\begin{eqnarray}
\Gamma(H_1\rightarrow {\psi}{\psi})& =\frac{M_{H_1} g_s^2 sin^2\theta}{8\pi} (1-\frac{4M_{{\psi}}^2}{{M_{H_1}^2}})^{3/2} .
\label{decayinv1}
\end{eqnarray}
 The SM prediction for the branching ratio of the Higgs boson decaying to invisible particles which coming from process $h\rightarrow ZZ^*\rightarrow 4\nu$ \cite{Denner:2011mq},\cite{Dittmaier:2012vm},\cite{Brein:2003wg},\cite{LHCHiggsCrossSectionWorkingGroup:2013rie} is, $1.2\times10^{-3}.$
CMS Collaboration has  reported the observed (expected) upper limit
on the invisible branching fraction of the Higgs boson to be $0.18 (0.10)$ at the $95\%$ confidence level, by assuming the SM production cross section \cite{CMS:2022qva}. A similar analysis was performed by ATLAS collaboration in which an observed upper limit of $0.145$ is placed on the branching fraction of its decay into invisible particles at a $95\%$ confidence level\cite{ATLAS:2022yvh}.

\begin{figure}%[!htb]
	\begin{center}
		\centerline{\hspace{0cm}\epsfig{figure=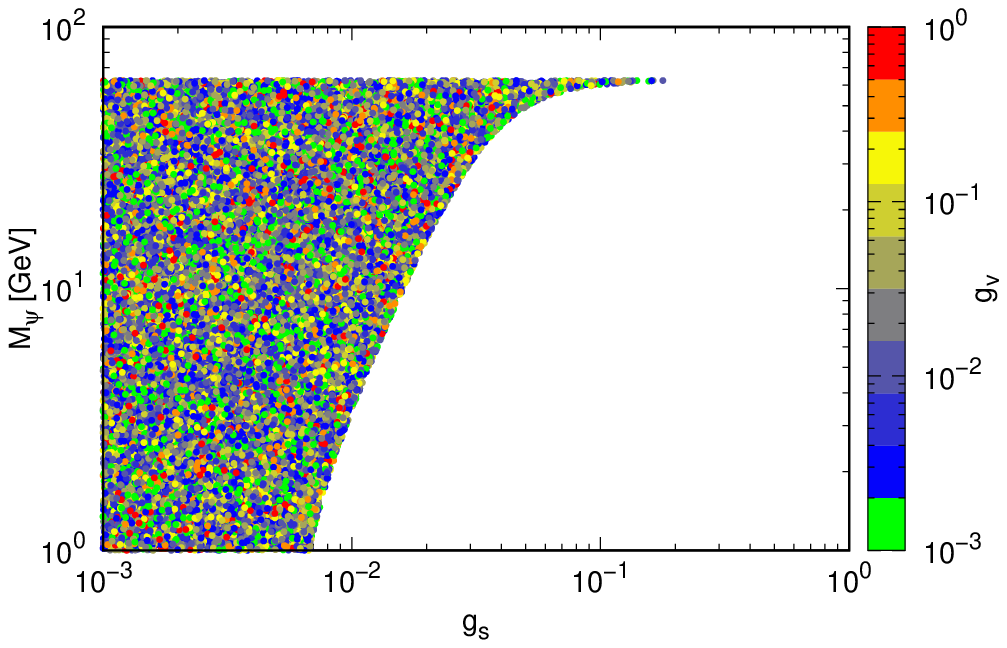,width=12cm}}
		\centerline{\vspace{-0.2cm}}
		\caption{ The cross points depict allowed region that is consistent with invisible Higgs decay at \cite{ATLAS:2022yvh}.
		} \label{Invisible}
	\end{center}
\end{figure}
Figure \ref{Invisible}, shows the allowed range of parameters by considering ATLAS \cite{ATLAS:2022yvh} upper limit for
invisible Higgs mode.

\subsection{Electroweak phase transition and gravitational waves}\label{phase transition}

\subsubsection{Finite temperature potential}
In addition to the 1-loop zero-temperature potential (\ref{2-16}), we can also consider the 1-loop corrections at finite
temperature in the effective potential, which is \cite{Dolan:1973qd}
\begin{equation}\label{finite}
V_{eff}^{1-loop(T\neq0)}(H_{2},T) = \frac{T^4}{2\pi^2}\sum_{k=1}^{n} g_{k} J_{B,F} \Bigl(\frac{M_k}{\nu}\frac{H_2}{T}\Bigl) ,
\end{equation}
with thermal functions
\begin{equation}
J_{B,F}(x)= \int_{0}^{\infty} dy  y^2  ln \Bigl(1\mp e^{-\sqrt{y^2+x^2}}\Bigl).
\end{equation}
The above functions can be expanded in terms of modified Bessel functions
of the second kind, $K_2 (x)$\cite{Mohamadnejad:2019vzg},
\begin{align}
& J_B (x)\simeq -\sum_{k=1}^{3} \frac{1}{k^2}x^2 K_2 (kx) , \nonumber \\
& J_F (x)\simeq -\sum_{k=1}^{2} \frac{(-1)^k}{k^2}x^2 K_2 (kx). \nonumber \\
\end{align}
The contribution of resummed daisy graphs is also as follows\cite{Carrington:1991hz}
\begin{equation}\label{daisy}
V_{daisy}(H_2 ,T)=  \sum_{k=1}^{n} \frac{g_k T^4}{12\pi}  \Biggl(\Bigl(\frac{M_k}{\nu}\frac{H_2}{T}\Bigl)^3 - \biggl(\Bigl(\frac{M_k}{\nu}\frac{H_2}{T}\Bigl)^2 + \frac{\Pi_k (T)}{T^2}\biggl)^{\frac{3}{2}} \Biggl),
\end{equation}
where the sum runs only over scalar bosons and longitudinal degrees of freedom of the
gauge bosons. Thermal masses, $\Pi_k (T)$, are given by
\begin{align}
& \Pi_W= \frac{11}{6}g_{SM}^2 T^2 \nonumber, ~~~~~~~~~~ \Pi_V=\frac{2}{3}g_{v}^2 T^2\nonumber,~~~~~~~~ \Pi_{Z/\gamma}=\frac{11}{6}\begin{pmatrix}g_{SM}^2~~~~~~0\\0~~~~~~g_{SM}^{\prime2}\end{pmatrix}T^2,\\
&  \Pi_{H_1 /H_2}=\begin{pmatrix}\frac{\lambda_H}{24}+\frac{\lambda_{SH}}{12}+\frac{3g_{SM}^2}{16}+\frac{g_{SM}^{\prime2}}{16}+\frac{\lambda_{t}^2}{4}~~~~~~~~~0~~~~~\\~~~~~~~~~~~~~~~~~0~~~~~~~~~~~~~~~~~~~\frac{\lambda_S}{24}+\frac{\lambda_{SH}}{12}+\frac{g_{v}^2}{4}\end{pmatrix}T^2.
\end{align}

Finally, the one-loop effective potential including both one-loop zero temperature \ref{2-16} and finite temperature \ref{finite} and \ref{daisy} corrections is given by
\begin{equation}\label{full potential}
V_{eff}(H_2 ,T)= V^{1-loop(T=0)}(H_2) +  V^{1-loop(T\neq0)}(H_2 ,T)+V_{daisy}(H_2 ,T)
\end{equation}

In order to get $V_{eff}(0 ,T)=0$ at all temperatures, We make the following substitution:
\begin{equation}
V_{eff}(H_2 ,T) \longrightarrow V_{eff}(H_2 ,T) - V_{eff}(0 ,T).
\end{equation}
Now we are ready to study the phase transition and the resulting gravitational waves.

\subsubsection{Gravitational waves}

The characteristic of first-order phase transitions is the existence of a barrier between the symmetric
and broken phases. The electroweak phase transition takes place after the
temperature of the universe drops below the critical
temperature($T_C$). At this temperature, effective potential(\ref{full potential}) has two
degenerate minimums, one in $H_2 =0$ and the other in $H_2=\nu_C \neq0$ :
\begin{align}
& V_{eff}(0 ,T_C)= V_{eff}(\nu_C ,T_C) , \nonumber \\
& \left. \frac{dV_{eff}(H_2 ,T_C)}{dH_2} \right|_{H_2=\nu_C}=0.
\end{align}
By solving these two equations, one can obtain $\nu_C$ and $T_C$ . If this phase transition is strongly first order, it can satisfy the condition of departure
from thermal equilibrium, which is one of Sakharov conditions for creating baryonic asymmetry in the universe. There is a criteria for strongly electroweak phase transition\cite{Shaposhnikov:1987tw,Shaposhnikov:1986jp} which is as follows
\begin{equation}
\frac{\nu_C}{T_C}>1.
\end{equation}

The transition from the false to the true vacuum proceeds via thermal tunneling
at finite temperature. This concept can be grasped in the context of formation of
bubbles of the broken phase in the sea of the symmetric phase. Once this has
happened, the bubble spreads throughout the universe, converting false vacuums
into true ones. Bubbles formation starts at the nucleation temperature $T_N$, where one can
 estimate $T_N$ by the condition $S_3 (T_N) / T_N \sim 140$\cite{Apreda:2001us}. The function $S_3(T)$ is the three-dimensional Euclidean
action for a spherical symmetric bubble given by
\begin{equation}\label{Euclidean action}
S_3(T)= 4\pi \int_{0}^{\infty} dr  r^2 \Biggl(\frac{1}{2} \Bigl(\frac{dH_2}{dr}\Bigl)^2 +V_{eff}(H_2 ,T)\Biggl),
\end{equation}
where $H_2$ satisfies the differential equation which minimizes $S_3$:
\begin{equation}\label{S3}
\frac{d^2 H_2}{dr^2} + \frac{2}{r} \frac{dH_2}{dr}=\frac{dV_{eff}(H_2 ,T)}{dH_2},
\end{equation}
with the boundary conditions:
\begin{equation}
\left. \frac{dH_2}{dr} \right|_{r=0}=0, ~~~~ and ~~~~ H_2 (r\longrightarrow \infty )=0.
\end{equation}
In order to solve Eq. \ref{S3} and find the Euclidean action \ref{Euclidean action}, we used the AnyBubble
package\cite{Masoumi:2017trx}. In the following, we will show that the nucleation temperature($T_N$) will be much lower than the critical temperature ($T_C$), indicating a very strong phase transition.

GWs resulting from the strong first-order electroweak phase transitions are have three causes, which are as follows:

$\bullet$ collisions of bubble walls and shocks in the plasma,

$\bullet$ sound waves to the stochastic background after collision of bubbles but before expansion

 has dissipated the kinetic energy in the plasmal,

 $\bullet$ turbulence forming after bubble collisions.

 These three processes may coexist, and each one contributes to the stochastic GW background:
\begin{equation}
\Omega_{GW} h^{2} \simeq \Omega_{coll} h^{2}+\Omega_{sw} h^{2}+\Omega_{turb} h^{2}.
\end{equation}

There are four thermal parameters that control the above contributions:

 $\bullet$ $T_N$:  the nucleation temperature,

  $\bullet$ $\alpha$: the ratio of the free energy density difference
between the true and false vacuum and

 the total energy density,
\begin{equation}
\alpha= \frac{\Delta \Bigl(V_{eff} -T\frac{\partial V_{eff}}{\partial T}\Bigl)\bigg\vert_{T_N}}{\rho_\ast},
\end{equation}

where $\rho_\ast$ is
\begin{equation}
\rho_\ast= \frac{\pi^2 g_\ast}{30}T_N^4,
\end{equation}

 $\bullet$ $\beta$:  the inverse time duration of the phase transition,
 \begin{equation}
\frac{\beta}{H_\ast}= T_N \frac{d}{dT}\Bigl(\frac{S_3 (T)}{T}\Bigl)\bigg\vert_{T_N},
\end{equation}

$\bullet$ $\upsilon_\omega$: the velocity of the bubble wall which is anticipated to be close to 1 for the strong transitions\cite{Bodeker:2009qy}.

Isolated spherical bubbles cannot be used as a source of
GWs, and these waves arise during the collision of the
bubbles. The collision contribution to the spectrum is
given by\cite{Huber:2008hg}
\begin{equation}
\Omega_{coll}(f) h^{2}=1.67\times 10^{-5} \Bigl(\frac{\beta}{H_\ast}\Bigl)^{-2} \Bigl(\frac{\kappa \alpha}{1+\alpha} \Bigl)^2 \Bigl(\frac{g_\ast}{100} \Bigl)^{-\frac{1}{3}} \Bigl(\frac{0.11 \upsilon_\omega^3}{0.42+\upsilon_\omega^2}\Bigl) S_{coll},
\end{equation}
where $S_{coll}$ parametrises the spectral shape and is given by
\begin{equation}
S_{coll}=\frac{3.8 (f/f_{coll})^{2.8}}{2.8 (f/f_{coll})^{3.8} +1 },
\end{equation}
where
\begin{equation}
f_{coll}= 1.65\times 10^{-5} \Bigl(\frac{0.62}{\upsilon_\omega^2 -0.1\upsilon_\omega +1.8}\Bigl)  \Bigl(\frac{\beta}{H_\ast}\Bigl) \Bigl(\frac{T_N}{100}\Bigl) \Bigl(\frac{g_\ast}{100} \Bigl)^{1/6} Hz.
\end{equation}

The collision of bubbles produces a massive movement in the fluid in the form of sound waves that generate GWs. This
is the dominant contribution to the GW signal, and is given by\cite{Hindmarsh:2015qta}
\begin{equation}
\Omega_{sw}(f) h^{2}=2.65\times 10^{-6} \Bigl(\frac{\beta}{H_\ast}\Bigl)^{-1} \Bigl(\frac{\kappa_\upsilon \alpha}{1+\alpha} \Bigl)^2 \Bigl(\frac{g_\ast}{100} \Bigl)^{-\frac{1}{3}} \upsilon_\omega S_{sw}.
\end{equation}
The spectral shape of $S_{sw}$ is
\begin{equation}
S_{sw}= (f/f_{sw})^3 \Bigl(\frac{7}{3 (f/f_{sw})^2 +4} \Bigl)^{3.5},
\end{equation}
where
\begin{equation}
f_{sw}= 1.9\times 10^{-5} \frac{1}{\upsilon_\omega} \Bigl(\frac{\beta}{H_\ast}\Bigl) \Bigl(\frac{T_N}{100}\Bigl) \Bigl(\frac{g_\ast}{100} \Bigl)^{1/6} Hz.
\end{equation}

Plasma turbulence can also be caused by bubble collisions, which is a contributing factor to the GW spectrum and is given by\cite{Caprini:2009yp}
\begin{equation}
\Omega_{turb}(f) h^{2}=3.35\times 10^{-4} \Bigl(\frac{\beta}{H_\ast}\Bigl)^{-1} \Bigl(\frac{\kappa_{turb} \alpha}{1+\alpha} \Bigl)^{3/2} \Bigl(\frac{g_\ast}{100} \Bigl)^{-\frac{1}{3}}  \upsilon_\omega S_{turb},
\end{equation}
where
\begin{equation}\label{Sturb}
S_{turb}= \frac{(f/f_{turb})^3}{(1+8\pi f/h_\ast)(1+f/f_{turb})^{11/3}},
\end{equation}
and
\begin{equation}
f_{turb}= 2.27\times 10^{-5} \frac{1}{\upsilon_\omega} \Bigl(\frac{\beta}{H_\ast}\Bigl) \Bigl(\frac{T_N}{100}\Bigl) \Bigl(\frac{g_\ast}{100} \Bigl)^{1/6} Hz.
\end{equation}

In Eq.\ref{Sturb}, $h_\ast$ is the value of the inverse Hubble time at
GW production, redshifted to today,
\begin{equation}
h_\ast= 1.65\times 10^{-5} \Bigl(\frac{T_N}{100}\Bigl) \Bigl(\frac{g_\ast}{100} \Bigl)^{1/6}.
\end{equation}

In computing GW spectrum we have used\cite{Caprini:2015zlo,Kamionkowski:1993fg}
\begin{align}
& \kappa= \frac{1}{1+0.715\alpha}(0.715\alpha + \frac{4}{27}\sqrt{\frac{3\alpha}{2}}) , \nonumber \\
& \kappa_\upsilon= \frac{\alpha}{0.73 + 0.083\sqrt{\alpha}+\alpha},~~~~ \kappa_{turb}=0.05\kappa_\upsilon ,
\end{align}
where the parameters $\kappa$, $\kappa_\upsilon$, and $\kappa_{turb}$ denote the fraction of latent heat that is transformed into gradient energy of the Higgs-like field, bulk motion of the fluid, and MHD turbulence, respectively.

\begin{figure}%[!htb]
	\begin{center}
		\centerline{\hspace{0cm}\epsfig{figure=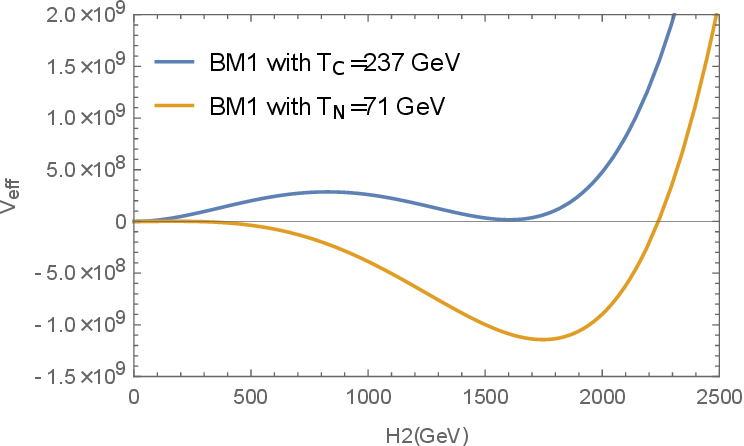,width=8cm}\hspace{0.5cm}\hspace{0cm}\epsfig{figure=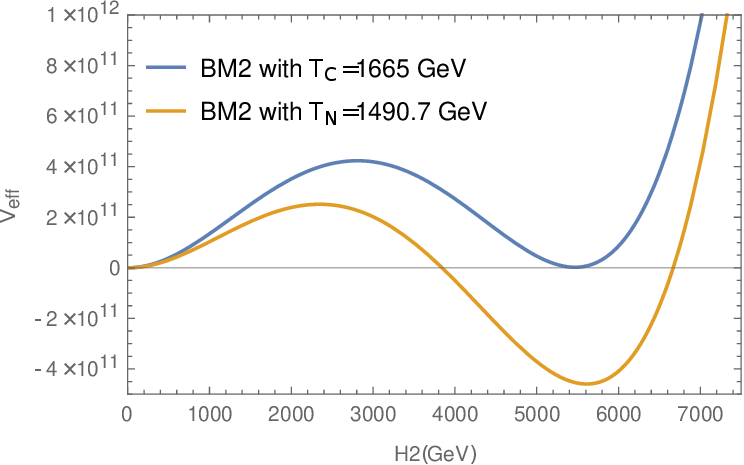,width=8cm}}
		\centerline{\vspace{0.2cm}\hspace{1.5cm}(a)\hspace{8cm}(b)}
		\centerline{\hspace{0cm}\epsfig{figure=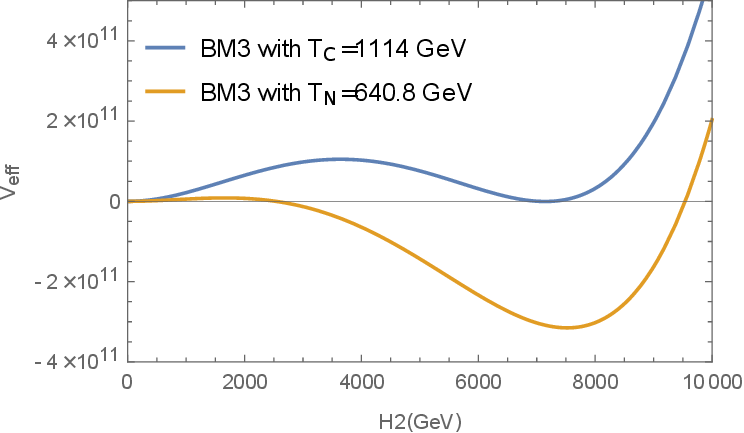,width=8cm}\hspace{0.5cm}\hspace{0cm}\epsfig{figure=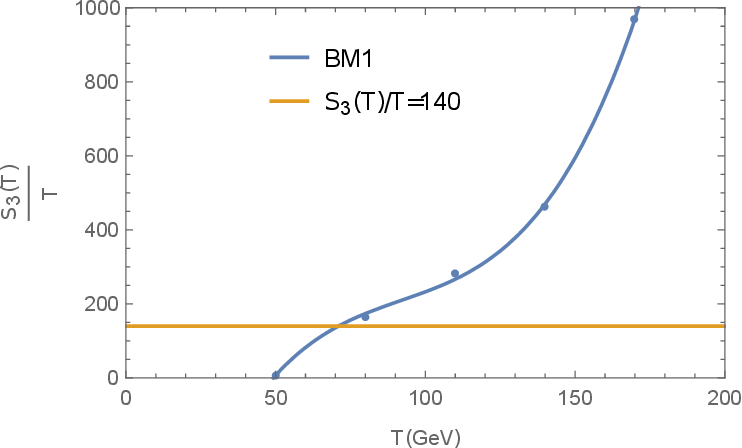,width=8cm}}		
		\centerline{\vspace{0.2cm}\hspace{1.5cm}(c)\hspace{8cm}(d)}
       \centerline{\hspace{0cm}\epsfig{figure=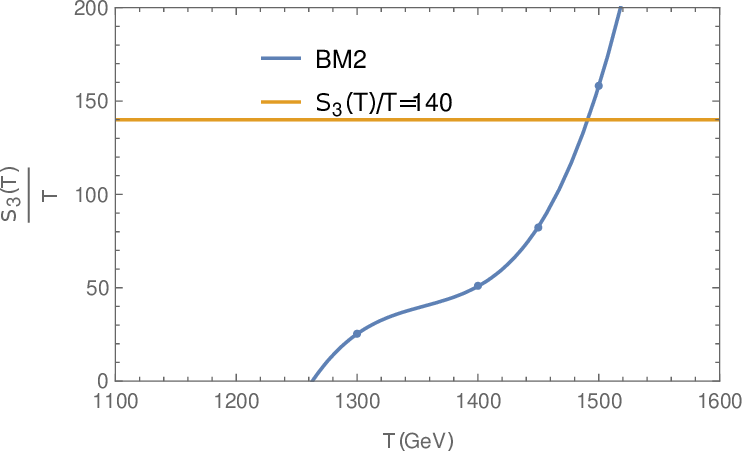,width=8cm}\hspace{0.5cm}\hspace{0cm}\epsfig{figure=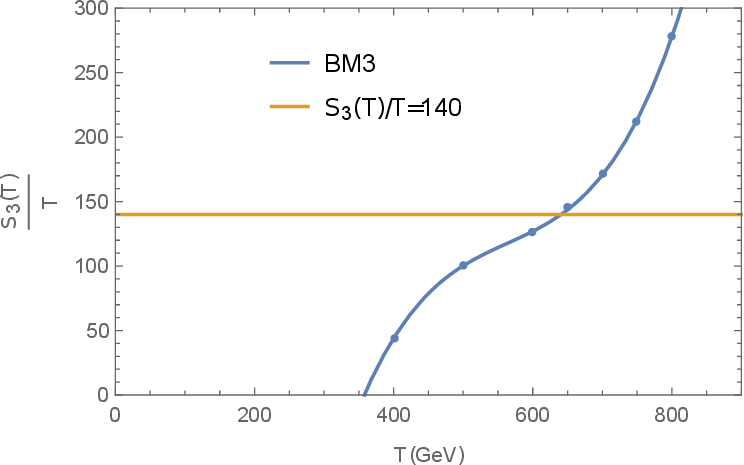,width=8cm}}
       \centerline{\vspace{0.2cm}\hspace{1.5cm}(e)\hspace{8cm}(f)}
		\centerline{\vspace{-0.2cm}}

		\caption{In (a), (b) and (c) Potential behavior are given for critical temperature and nucleation temperature. In (d), (e) and (f) $S_3/T$ changes in terms of temperature are also given for all three benchmarks.} \label{S3/T2}
	\end{center}
\end{figure}

\begin{figure}%[!htb]
	\begin{center}
		\centerline{\hspace{0cm}\epsfig{figure=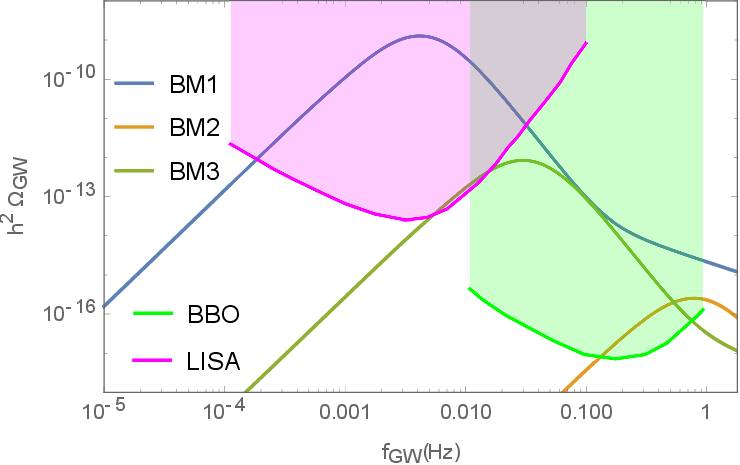,width=12cm}}
		\centerline{\vspace{-0.2cm}}
		\caption{GW spectrum for benchmark points of the Table \ref{table2}.} \label{GW2}
	\end{center}
\end{figure}

To investigate the GWs resulting from the first-order  electroweak phase transition, we choose three benchmark points.  These points are presented in table \ref{table2}. Figure \ref{S3/T2} shows the potential behavior for both critical and nucleation temperatures. Also, $S_3/T$ changes in terms of temperature are shown. In table \ref{table2} all relevant quantities, including
independent parameters of the model, DM properties, and phase transition parameters,
are given. The benchmarks 1 and 2 are consistent with direct detection constraint while the benchmark 3 is outside the range of XENONnT and is placed under the neutrino floor. The GW spectrum for these benchmark points is depicted in figure \ref{GW2}. The GW spectrum for these benchmarks 1,(2,3) falls within the observational window of LISA(BBO). Therefore, for benchmark 3, GW can be a special way to probe it.

\begin{table}[h]
\centering % centering table
\begin{tabular}{l c c rrrrrrr} % creating 10 columns
\hline\hline
 $\#$ &$M_V (GeV)$ &$M_{\psi}(GeV)$ &$g_v$&$g_s$&$M_{H_2}(GeV)$&$\Omega_{{\psi}} h^2 $&$\sigma_{{\psi}} (cm^2)$\\
\hline
1&839.9&36.1&0.485&0.029&78.39&$1.11\times10^{-1}$&$1.11\times10^{-47}$ \\
2&4998&407.8&0.87&0.1&849.1&$1.18\times10^{-1}$&$5.16\times10^{-48}$\\
3&3526&159.7&0.468&0.029&321.6&$1.2\times10^{-1}$&$1.98\times10^{-49}$ \\
\hline\hline
$\#$ &$T_C (GeV)$&$T_N (GeV)$&$\alpha$&$\beta/H_\ast$&$(\Omega_{GW} h^2)_{max}$\\
\hline
1&237&71&1.48&306.694&$1.32\times10^{-9}$&-----&-----\\
2&1665&1490.7&0.02&2758.95&$2.68\times10^{-16}$&-----&-----\\
3&1114&640.8&0.09&244.78&$8.77\times10^{-13}$&-----&-----\\
\hline
\end{tabular}
\caption{\label{table2}Three benchmark points with DM and phase transition parameters.} % title name of the table
\end{table}

\section{Scenario B}\label{Scenario b}

\subsection{DM phenomenology}

\subsubsection{Relic density}
\begin{figure}%[!htb]
	\begin{center}
		\centerline{\hspace{0cm}\epsfig{figure=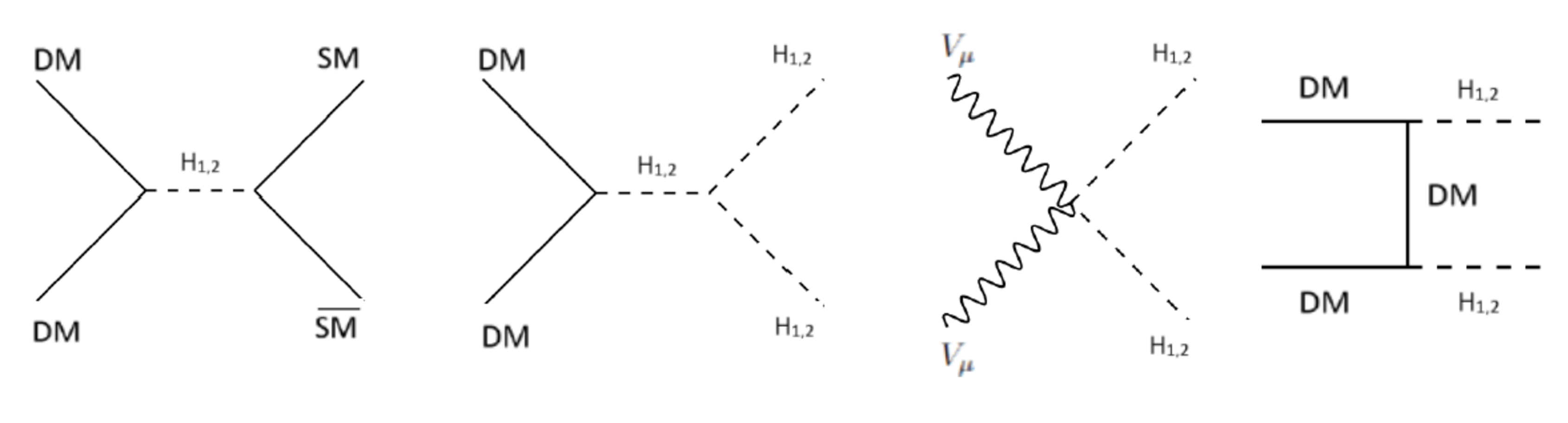,width=10cm}\hspace{0.5cm}\hspace{0cm}}
		\centerline{\vspace{0.2cm}\hspace{1.5cm}(a)}
       \centerline{\hspace{0cm}\epsfig{figure=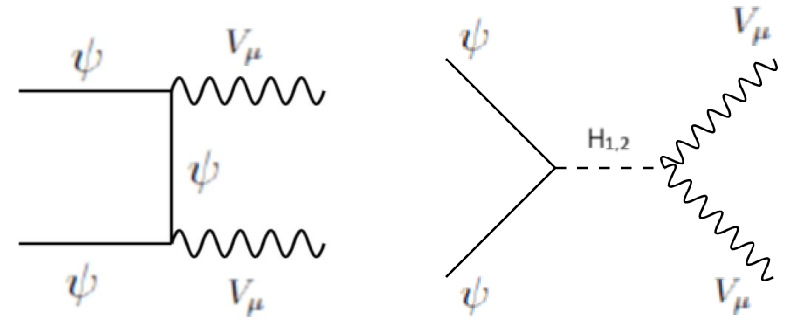,width=6cm}\hspace{0.5cm}\hspace{0cm}}		
		\centerline{\vspace{0.2cm}\hspace{1.5cm}(b)}
		\centerline{\hspace{0cm}\epsfig{figure=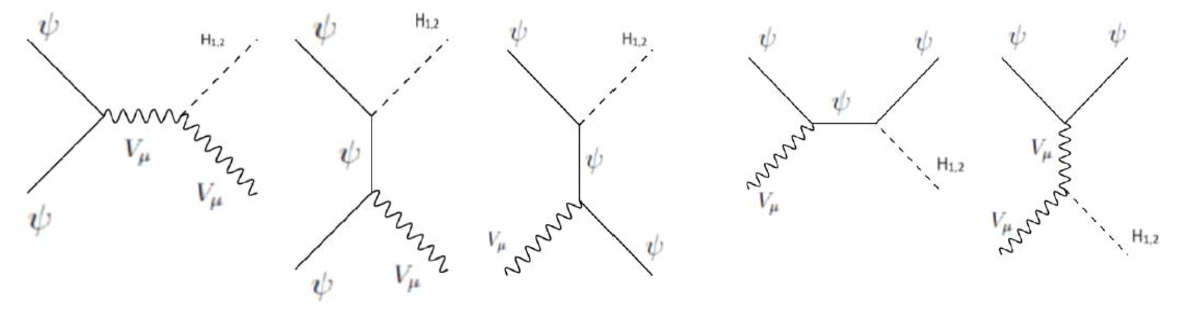,width=12cm}\hspace{0.5cm}\hspace{0cm}}		
		\centerline{\vspace{0.2cm}\hspace{1.5cm}(c)}

		\caption{The relevant Feynman diagrams for DM relic density cross section including: a(annihilation), b(conversion) and c(semi-annihilation).} \label{feynman}
	\end{center}
\end{figure}

 In this scenario, $M_V< 2M_{\psi} $ and both $V_{\mu}$ and $\psi$ fields are considered as DM. The evolution of the number density of DM particles with time are governed by the Boltzmann equation. The coupled Boltzmann equations for fermion ${\psi}$ and vector DM are given by:
\begin{equation} \label{44}
\frac{dn_V}{dt}+3Hn_V = -\sum_{j} \langle \sigma_{VV \rightarrow jj}\upsilon\rangle (n_V ^2 -n_{V,eq} ^2)-\langle \sigma_{VV \rightarrow{\psi}{\psi} }\upsilon\rangle (n_V ^2-n_{V,eq} ^2 \frac{n_{\psi} ^2}{n_{{\psi},eq} ^2}) ,
\end{equation}
\begin{equation} \label{45}
\frac{dn_{\psi}}{dt}+3Hn_{\psi} = -\sum_{j} \langle \sigma_{{\psi}{\psi} \rightarrow jj}\upsilon\rangle (n_{\psi} ^2 -n_{{\psi},eq} ^2)-\langle \sigma_{{\psi}{\psi} \rightarrow VV }\upsilon\rangle (n_{\psi} ^2-n_{{\psi},eq} ^2 \frac{n_V ^2}{n_{V,eq} ^2}) ,
\end{equation}
where $j$ runs over SM massive particles, $H_1$ and $H_2$. In $\langle \sigma_{ab \rightarrow cd}\upsilon\rangle$ all annihilations are taken into account except $\langle \sigma_{{\psi}V \rightarrow {\psi}V}\upsilon\rangle$ which does not affect the number density. In the above relations, for simplicity in the writing, only the annihilation and conversion contributions are shown in the equations. But in practice, all contributions, even  semi-annihilations are included in the micrOMEGAs package to solve these equations\cite{Belanger:2014vza}. The relevant Feynman diagrams for DM production are shown in Fig~\ref{feynman}. By choosing $x=M/T$ and $Y=n/s$, , where $T$ and $s$ are the photon temperature and the entropy density, respectively, one can
rewrite the Boltzmann equations in terms of  $Y = n/s$:
\begin{equation} \label{46}
\frac{dY_V}{dx}=-\sqrt{\frac{45}{\pi}} M_{pl}  g_* ^{1/2} \frac{M}{x^2} [\sum_{j} \langle \sigma_{VV \rightarrow jj}\upsilon\rangle (Y_V ^2 -Y_{V,eq} ^2)+\langle \sigma_{VV \rightarrow{\psi}{\psi} }\upsilon\rangle (Y_V ^2-Y_{V,eq} ^2 \frac{Y_{\psi} ^2}{Y_{{\psi},eq} ^2})],
\end{equation}
\begin{equation} \label{47}
\frac{dY_{\psi}}{dx}=-\sqrt{\frac{45}{\pi}} M_{pl}  g_* ^{1/2} \frac{M}{x^2}[\sum_{j} \langle \sigma_{{\psi}{\psi} \rightarrow jj}\upsilon\rangle (Y_{\psi} ^2 -Y_{{\psi},eq} ^2)+\langle \sigma_{{\psi}{\psi} \rightarrow VV }\upsilon\rangle (Y_{\psi} ^2-Y_{{\psi},eq} ^2 \frac{Y_V ^2}{Y_{V,eq} ^2})],
\end{equation}
where $g_* ^{1/2}$ is the degrees of freedom parameter and $M_{pl}$ is the Planck mass. It is clear from the above equations that there are new terms in the Boltzmann equations that describe the conversion of two DM particles into each other. Because these two cross sections are also described by the same matrix
element, we expect $\langle \sigma_{VV \rightarrow{\psi}{\psi} }\upsilon\rangle$  and $\langle \sigma_{{\psi}{\psi} \rightarrow VV }\upsilon\rangle$ are not independent and their relation is:
\begin{equation}
Y_{V,eq} ^2 \langle \sigma_{VV \rightarrow{\psi}{\psi} }\upsilon\rangle=Y_{{\psi},eq} ^2 \langle \sigma_{{\psi}{\psi} \rightarrow VV }\upsilon\rangle .
\end{equation}

The interactions between the two DM components take place by exchanging two scalar mass eigenstates $H_1$ and $H_2$ where the coupling of $V$ to $H_1$ is suppressed by $\sin\alpha$. For this reason, it usually is the $H_2$ -mediated diagram that gives the dominant contribution. We also know that
the conversion of the heavier particle into the lighter one is relevant. The relic density for any DM candidate associated with the $Y$ at the present temperature is given by the following relation:
\begin{equation}
\Omega_{{\psi},V}h^2 = 2.755\times 10^8 \frac{M_{{\psi},V}}{GeV} Y_{{\psi},V}(T_0)
\end{equation}
where $h$ is the Hubble expansion rate at present time in units
of $ 100 (km/s)/Mpc$. We used the micrOMEGAs package\cite{Belanger:2014vza} to numerically solve coupled Boltzmann differential equations. According to the data from the Planck Collaboration\cite{Planck:2018vyg}, the DM
constraint in this model reads
\begin{equation}
\Omega_{DM} h^{2} = \Omega_{V} h^{2}+ \Omega_{{\psi}} h^{2} = 0.120 \pm 0.001.
\end{equation}
We also define the fraction of the DM density of each component by,
\begin{equation}
\xi_V= \frac{\Omega_V}{\Omega_{DM}}, ~~~~~ \xi_{\psi}= \frac{\Omega_{{\psi}}}{\Omega_{DM}}, ~~~ \xi_V+\xi_{\psi} =1.
\end{equation}

\begin{figure}%[!htb]
	\begin{center}
		\centerline{\hspace{0cm}\epsfig{figure=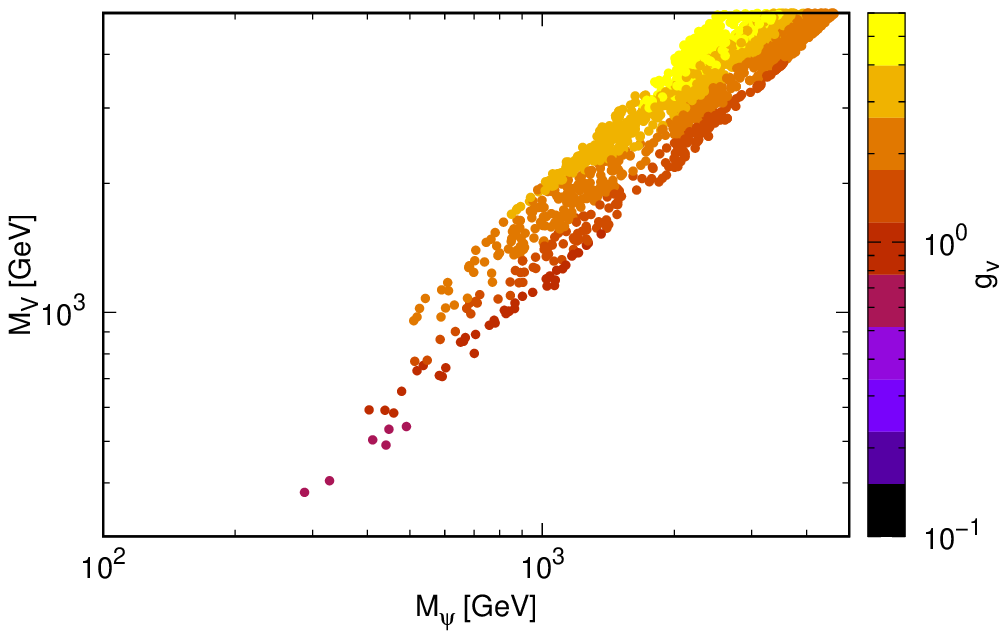,width=12cm}}
		\centerline{\vspace{-0.2cm}}
		\caption{The allowed range of parameter space consistent with DM relic density.} \label{Relic}
	\end{center}
\end{figure}

In Fig.~\ref{Relic}, the parameter space consistent with DM relic density is
obtained. As can be seen, there is an agreement with the relic density observed for $300< M_V<5000$ GeV, $200<M_{\psi}<5000$ GeV and $0.1<  g_v <6$. Of course, it is necessary to mention that the contribution of semi-annihilations in the model is important. For example, for benchmark 2 in Table \ref{table}, using the micrOMEGAs package, we found that the share of the cross section of the process of V $\psi$ $\to$ $\psi$ $H_{2}$ includes 24 \% of the cross section of all processes.

\subsubsection{Direct detection}
 We investigate constraints on parameters space of the model which are imposed by searching for scattering of DM-nuclei. The spin-independent direct detection(DD) cross sections of $V$ and ${\psi}$ are determined by $H_1$ and $H_2$ exchanged
diagrams\cite{YaserAyazi:2019caf,YaserAyazi:2018lrv}:
\begin{equation}
\sigma_{DM-N} ^V =\xi_{V} \frac{4 \lambda_{SH} ^2 M_V ^2 M_N ^2 \mu_{VN} ^2 (M_{H1} ^2 -M_{H2} ^2 )^2}{\pi M_{H1} ^8 M_{H2} ^4}f_N ^2 ,
\end{equation}
\begin{equation}\label{direct formula}
\sigma_{DM-N} ^{\psi} = \xi_{\psi} \frac{ g_s ^3 \nu_1 }{\pi M_{\psi} (1+(\nu_1 g_s/M_{\psi})^2 )} \mu_{\psi} ^2 (\frac{1}{M_{H1} ^2}-\frac{1}{M_{H2} ^2})^2 f_N ^2  ,
\end{equation}
where
\begin{equation}
\mu_{VN} = M_N M_V / (M_N + M_V) , ~~~~~ \mu_{\psi}= M_N M_{\psi} / (M_N + M_{\psi}).
\end{equation}
$M_N$ is the nucleon mass and $f_N\simeq 0.3$ parameterizes the Higgs-nucleon coupling.

Various DD  experiments have placed constraints on DM-Nucleon spin
independent cross section, such as LUX\cite{LUX:2016ggv}, PandaX-II\cite{PandaX-II:2016vec}, XENON1T\cite{XENON:2018voc},LZ\cite{LZ:2022ufs} and XENONnT\cite{XENON:2023sxq}. Of course, these experiments are gradually approaching what is called the neutrino floor\cite{Billard:2013qya}, which is a the irreducible background coming from scattering of SM neutrinos on nucleons. We use the XENONnT\cite{XENON:2023sxq} experiment results to constrain the parameter space of our model. In this experiment there is a minimum upper
limit on the spin-independent WIMP-nucleon cross section of $2.58 \times 10^{-47}$ $cm^2$ for a WIMP mass of
$28$ GeV . In order to study
the effect of the direct detection experiment on the model, we use rescaled DM-Nucleon cross section $\xi_V \sigma_V $ and $\xi_{\psi} \sigma_{\psi} $.

\begin{figure}%[!htb]
	\begin{center}
		\centerline{\hspace{0cm}\epsfig{figure=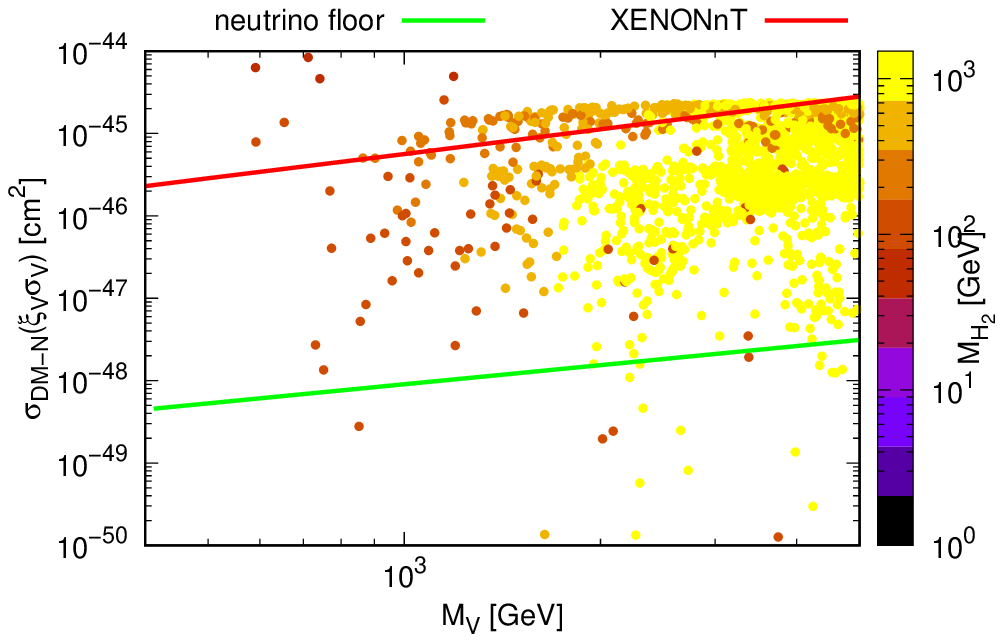,width=8cm}\hspace{0.5cm}\hspace{0cm}\epsfig{figure=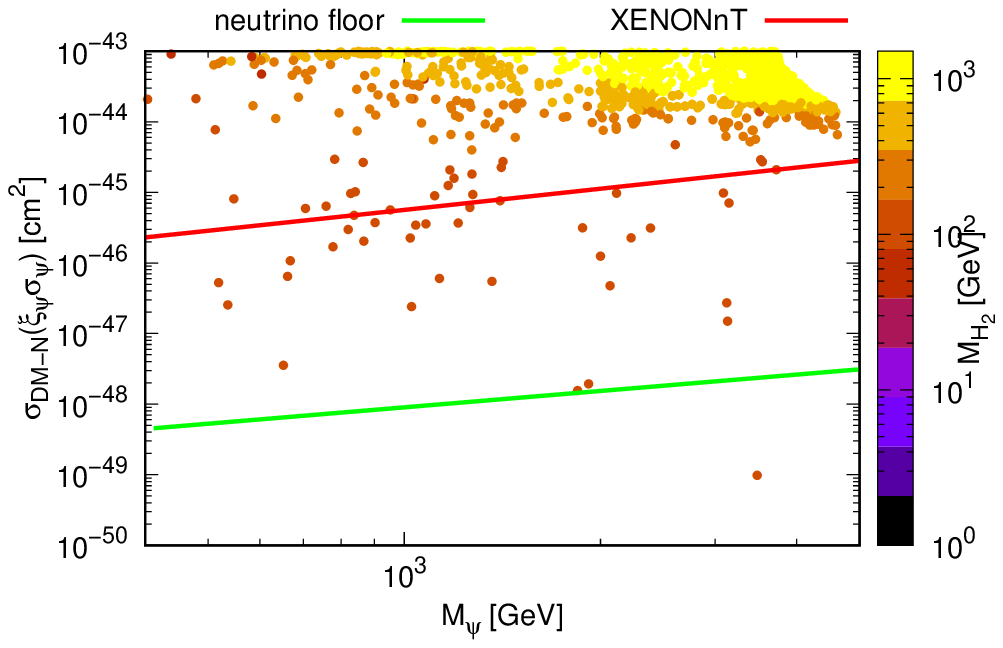,width=8cm}}
		\centerline{\vspace{0.2cm}\hspace{1.5cm}(a)\hspace{8cm}(b)}
\caption{The allowed range of parameter space consistent with DM relic density and DD.  In (a) $\xi_V \sigma_V$ VS $M_V$ and in (b) $\xi_{\psi} \sigma_{\psi}$ VS $M_{\psi}$ has shown.} \label{direct detection}
	\end{center}
\end{figure}

In Fig.~\ref{direct detection}, rescaled DM-Nucleon cross sections($\xi_V \sigma_V $ and $\xi_{\psi} \sigma_{\psi} $) are depicted for the parameters that are in agreement with the relic density. It is clear from the figure that there are some
points between the XENONnT direct detection bound and the neutrino floor which can
be probed in future direct detection experiments. As is expected from \ref{direct formula}, by reducing the mass difference between $H_1$ and $H_2$, the cross section decreases and therefore allowable points increase. In our model, DM interacts with nucleons through $H_{1}$ and $H_{2}$ mediators. The relevant terms in the Lagrangian are $A h_{1}[\bar{q} q]$ and $B h_{2} [ V_{\mu} V^{\mu}$ $ (\bar{\psi} \psi)] $ for vector (spinor) DM, where $ A $ and $ B $ are some constants.
Therefore, the $H_{1}$ mediator involves $A \cos \alpha H_{1} [\bar{q} q]-B\sin \alpha H_{1} [ V_{\mu} V^{\mu}$ $ (\bar{\psi} \psi)]$. Similarly, for the $H_{2}$ mediator, the terms are $A \sin \alpha H_{2} [\bar{q} q]+B\cos \alpha H_{2} [ V_{\mu} V^{\mu}$ $ (\bar{\psi} \psi)]$. Consequently, the effective 5(6)-dimensional interaction terms for DM-quark interactions at low energies will be $AB \sin \alpha \cos \alpha\left(\frac{1}{M_{H_{2}}^{2}}-\frac{1}{M_{H_{1}}^{2}}\right) [\bar{q} q] [ V_{\mu} V^{\mu}$ $ (\bar{\psi} \psi)]$. Around $M_{H_{2}} \simeq M_{H_{1}}$, the effective coupling between DM and quarks approaches zero, resulting in a dip in the DM-nucleon cross-section. In \cite{Abe:2021nih}, a study has been done on degenerate Higgs scenario. In \cite{CMS:2014afl}, it is shown by using the high resolution of the diphoton channel
	of the Higgs boson decays, the mass difference between the two degenerate states $\Delta m \gtrsim 3~ \rm GeV$	is disfavored at the $2\sigma$ level from the LHC Run-I data. In order to test a degenerate Higgs scenario, a possible proposal is consideration of degenerate scalar productions at the  International Linear Collider (ILC)\cite{Behnke:2013xla}. It was shown\cite{Abe:2021nih} that for  $\Delta m \gtrsim 1~\rm GeV$, it is possible to distinguish between two Higgs.

\subsubsection{Invisible Higgs decay}
In this scenario, according to Fig.~\ref{direct detection}, because there is no point where $M_{\psi,V,H_2}<M_{H1}/2$, it is not necessary to check the invisible Higgs decay.

\subsection{Electroweak phase transition and gravitational waves}

\begin{figure}%[!htb]
	\begin{center}
		\centerline{\hspace{0cm}\epsfig{figure=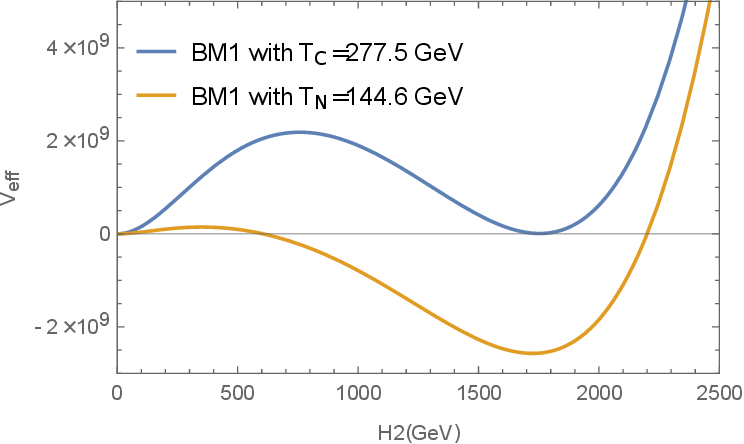,width=8cm}\hspace{0.5cm}\hspace{0cm}\epsfig{figure=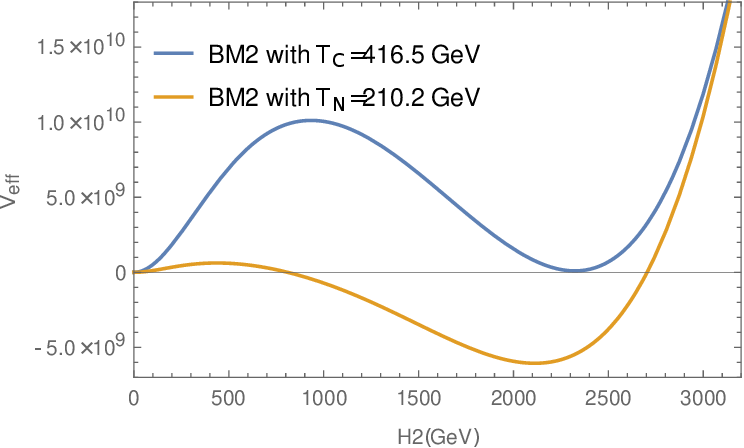,width=8cm}}
		\centerline{\vspace{0.2cm}\hspace{1.5cm}(a)\hspace{8cm}(b)}
		\centerline{\hspace{0cm}\epsfig{figure=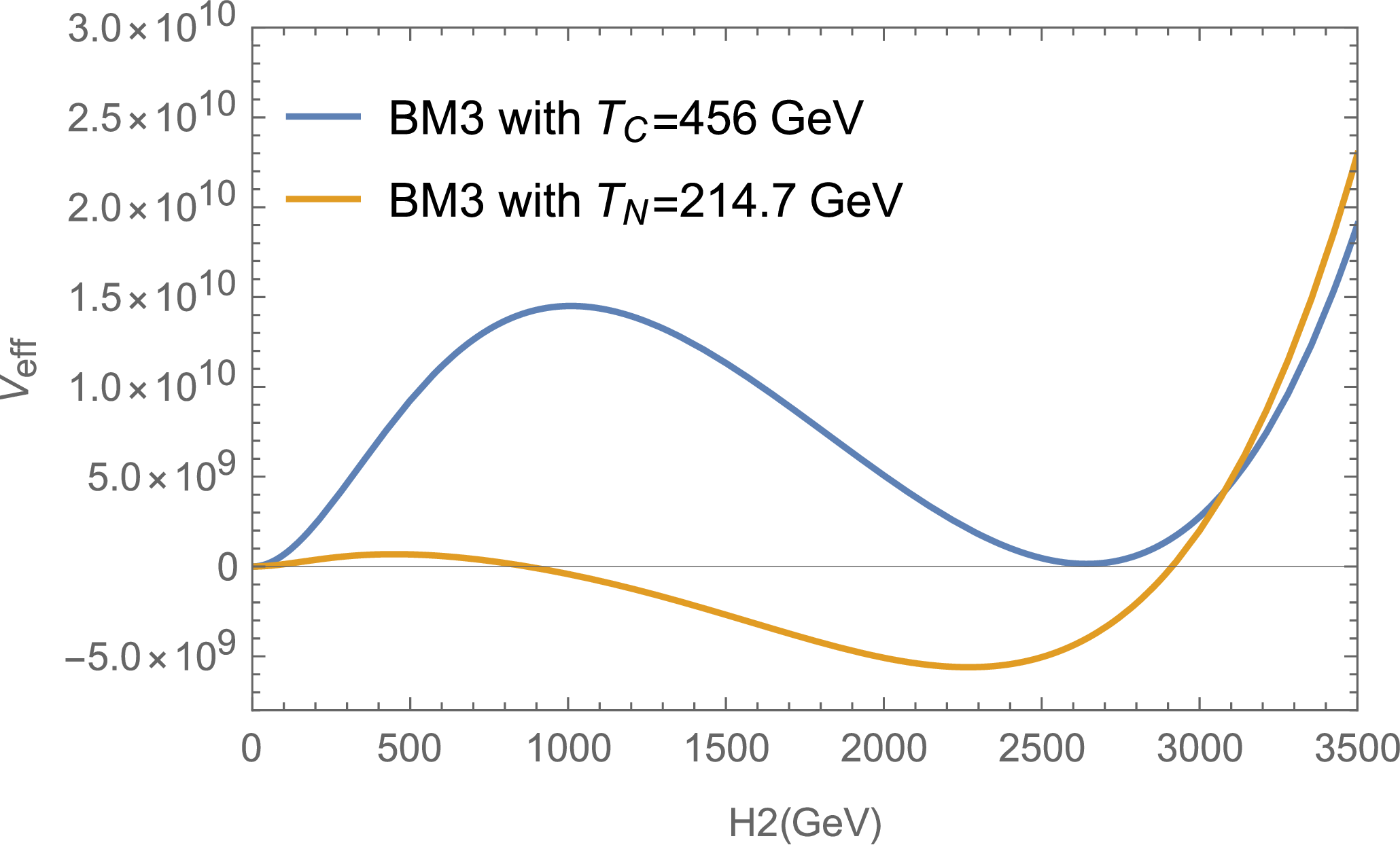,width=8cm}\hspace{0.5cm}\hspace{0cm}\epsfig{figure=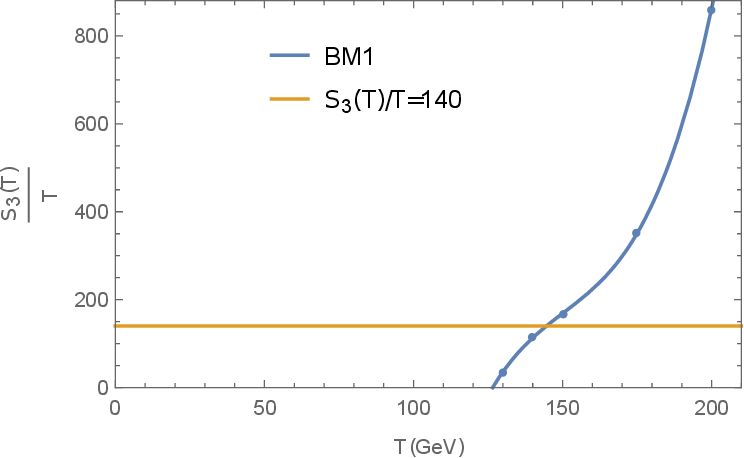,width=8cm}}		
		\centerline{\vspace{0.2cm}\hspace{1.5cm}(c)\hspace{8cm}(d)}
       \centerline{\hspace{0cm}\epsfig{figure=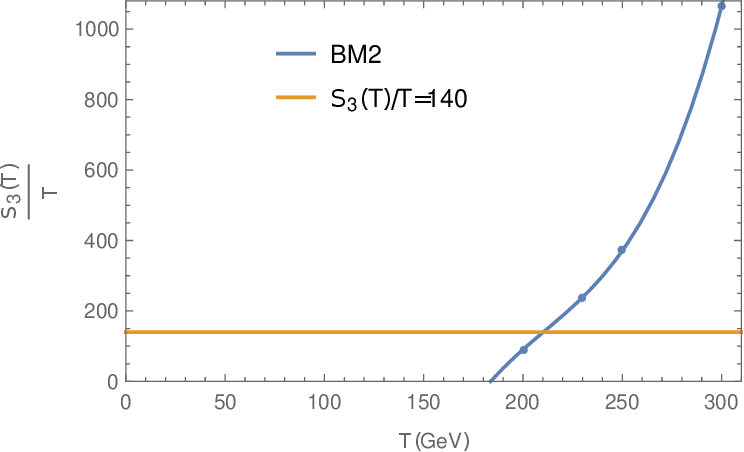,width=8cm}\hspace{0.5cm}\hspace{0cm}\epsfig{figure=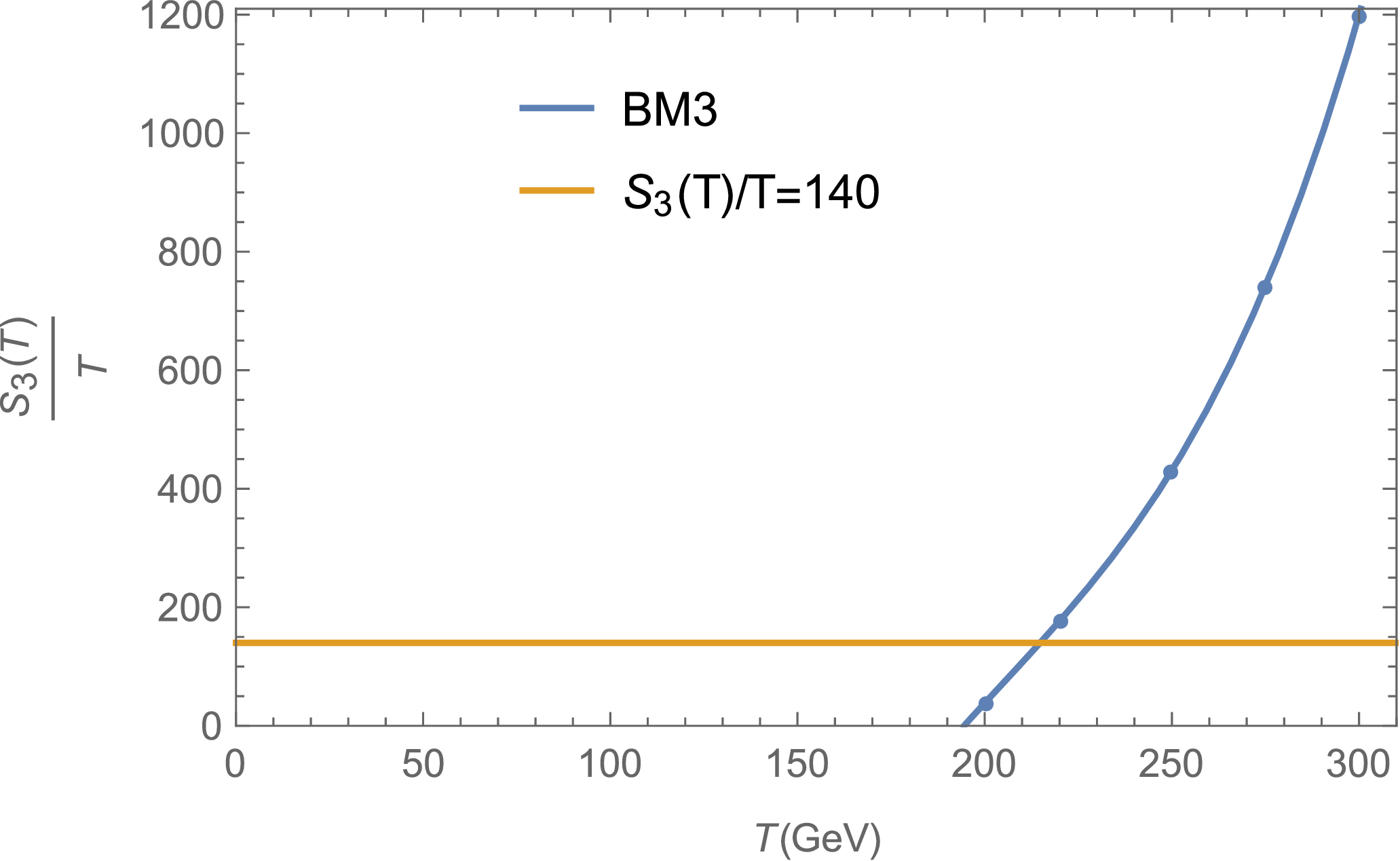,width=8cm}}
       \centerline{\vspace{0.2cm}\hspace{1.5cm}(e)\hspace{8cm}(f)}
		\centerline{\vspace{-0.2cm}}

		\caption{In (a), (b) and (c) Potential behavior are given for critical temperature and nucleation temperature. In (d), (e) and (f) $S_3/T$ changes in terms of temperature are also given for all three benchmarks.} \label{S3/T}
	\end{center}
\end{figure}

\begin{figure}%[!htb]
	\begin{center}
		\centerline{\hspace{0cm}\epsfig{figure=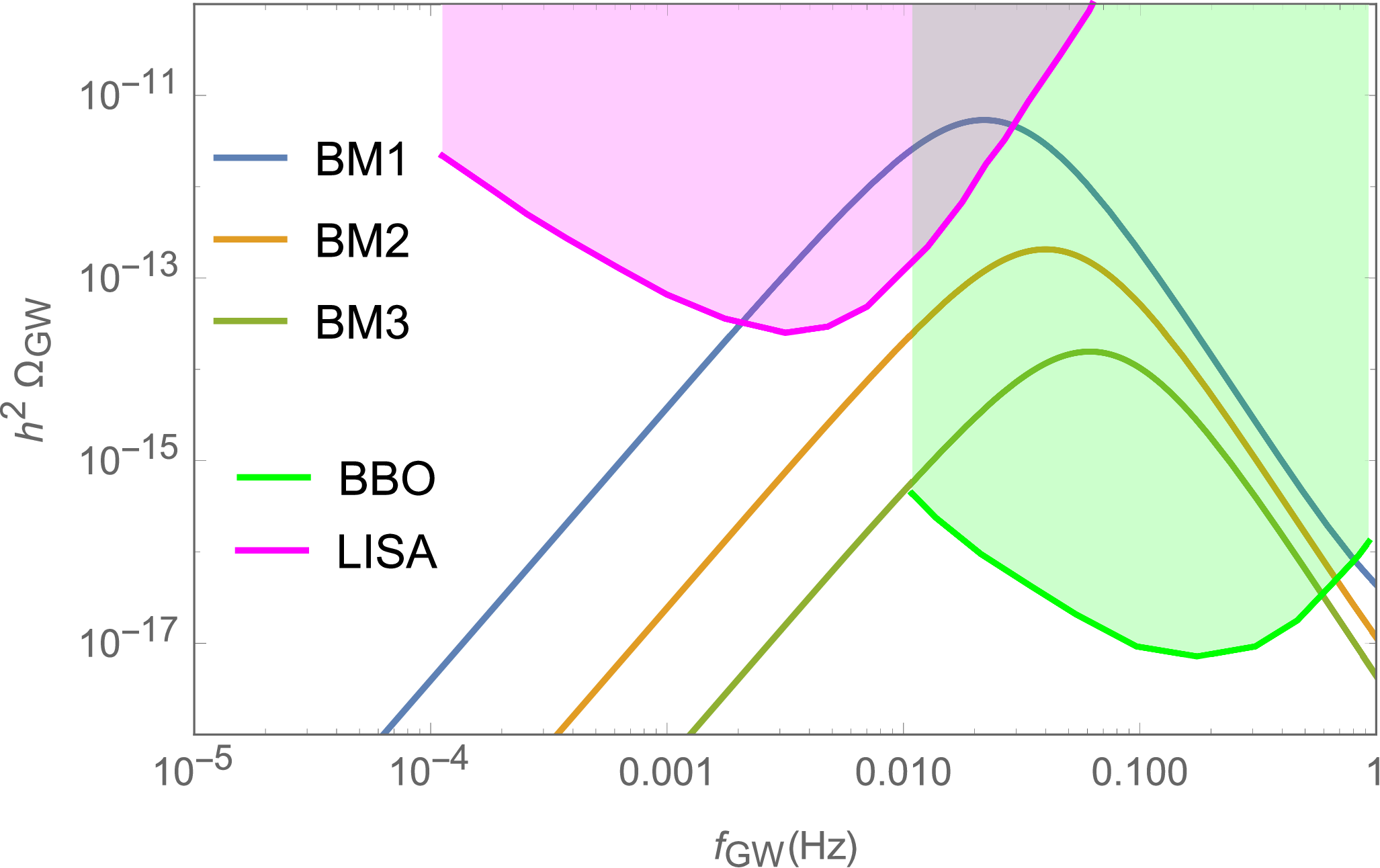,width=12cm}}
		\centerline{\vspace{-0.2cm}}
		\caption{GW spectrum for benchmark points of the Table \ref{table}.} \label{GW}
	\end{center}
\end{figure}

To investigate the phase transition and the resulting GWs, we follow the procedure of Sec.~\ref{phase transition}. We select three benchmark points as shown in Table \ref{table}. Benchmarks 1 and 2 are consistent with direct detection constraint, while benchmark 3 is placed under the neutrino floor. Figure \ref{S3/T} shows the changes of potential and $S_3/T$ in terms of temperature. The GW spectrum for these benchmark points is depicted in Fig.~\ref{GW}. The GW spectrum for these benchmarks falls within the observational window of BBO. For benchmark 1, gravitational wave peak will also be in LISA's range. For the benchmark 3, which is below the neutrino floor limit, future gravitational wave detectors can provide a special way to discover and investigate.

\begin{table}[h]
\centering % centering table
\begin{tabular}{l c c rrrrrrr} % creating 10 columns
\hline\hline
 $\#$ &$M_V (GeV)$ &$M_{\psi}(GeV)$ &$g_v$&$g_s$&$M_{H_2}(GeV)$\\
\hline
1&2249&2069&1.32&1.72&121.7 \\
2&3400&3150&1.65&2.16&140\\
3&3750&3481&1.71&2.25&125.1 \\
\hline\hline % inserts single-line
$\#$ &$\Omega_{V} h^{2}$ &$\Omega_{{\psi}} h^{2}$ &$\Omega_{DM} h^2 $&$\xi_V \sigma_V (cm^2)$&$\xi_{{\psi}}\sigma_{{\psi}} (cm^2)$\\
\hline
1&$1.17\times10^{-2}$&$1.09\times10^{-1}$&$1.2\times10^{-1}$&$6.02\times10^{-48}$&$4.75\times10^{-47}$\\
2&$1.1\times10^{-2}$&$9.99\times10^{-2}$&$1.1\times10^{-1}$&$9.03\times10^{-47}$&$7.04\times10^{-46}$\\
3&$1.15\times10^{-2}$&$1.04\times10^{-1}$&$1.15\times10^{-1}$&$1.26\times10^{-50}$&$9.84\times10^{-50}$\\
\hline\hline
$\#$ &$T_C (GeV)$&$T_N (GeV)$&$\alpha$&$\beta/H_\ast$&$(\Omega_{GW} h^2)_{max}$\\
\hline
1&277.5&144.6&0.22&796.068&$5.26\times10^{-12}$\\
2&416.5&210.2&0.09&996.56&$2.08\times10^{-13}$\\
3&456&214.7&0.05&1501.05&$1.53\times10^{-14}$\\
\hline
\end{tabular}
\caption{\label{table}Three benchmark points with DM and phase transition parameters.} % title name of the table
\end{table}

\section{Conclusion} \label{Conclusion}
We have considered an extension of the SM  with a new $U(1)$ symmetry in the dark part. According to the new particle mass in the model, we considered two different scenarios. The model consists of three new fields: a fermion, a complex scalar, and a vector field. In scenario A, only fermionic particles are considered as DM. In scenario B, both vector and fermionic particles are considered as DM. The model has classical scale symmetry; electroweak symmetry breaking occurs through Gildener–
Weinberg mechanism and gives a natural solution to the hierarchy problem. Numerical solution of the Boltzmann equations for both scenarios was conducted to determine a parameter space region which is compatible with Planck and XENONnT data  and collider constraints (invisible Higgs decay in scenario A).A three-dimensional parameter space acquisition was completed.

We focused our attention on the phase transition dynamics after presenting the model and exploring DM phenomenology. The full finite-temperature effective potential of the model at the one-loop level was obtained to investigate the nature and strength of the electroweak phase transition, with the aim of exploring its nature and strength.
A first-order electroweak phase transition can exist when there is a barrier between the broken and symmetric phases. It was demonstrated that the finite-temperature effects induce such a barrier and thereby give rise to a phase transition which can generate GWs.

After studying the phase transition, we investigated the resulting GWs. The parameters required to investigate GWs can all be calculated from our presented model and are a function of the independent parameters of the model. We have demonstrated that the model can survive DM relic density, direct detection and collider constraints, while also producing GWs during the first order electroweak phase transition. We also showed that GWs can be a special probe for the benchmarks that are placed under the neutrino floor(benchmark 3 in both scenarios). These waves can be placed within the observation window of LISA and BBO, which is hoped to be a path to new physics.

\bibliography{References}
\bibliographystyle{JHEP}

\end{document}